\documentclass[amsmath,amssymb]{nature}

\usepackage{ifthen}		
\newboolean{SubmittedVersion}
\setboolean{SubmittedVersion}{False} 	

\usepackage{graphicx}
\usepackage{amsmath}
\usepackage{amssymb}
\DeclareMathAlphabet{\mathpzc}{OT1}{pzc}{m}{it}

\def\nm{{\ \text{nm}}}                       
\def\cm{{\ \text{cm}}}                       
\def\micron{{\ \mu\text{m}}}                 

\def\Hz{{\ \text{Hz}}}                       
\def\kHz{{\ \text{kHz}}}                     
\def\MHz{{\ \text{MHz}}}                     

\def\us{{\ \mu\text{s}}}                     
\def\ms{{\ \text{ms}}}                       
\def\second{{\ \text{s}}}                    

\def\kl{k_L}                            
\def\Rb87{^{87}\text{Rb}}                     
\def\Na23{^{23}\text{Na}}                     
\def\Li6{^{6}\text{Li}}                       

\def\ket#1{\mathinner{|{#1}\rangle}}

{\catcode`\|=\active
  \gdef\Braket#1{\left<\mathcode`\|"8000\let|\BraVert {#1}\right>}}
\def\BraVert{\egroup\,\mid@vertical\,\bgroup}

\title{Spin-orbit-coupled Bose-Einstein condensates}
\author{Y.-J.~Lin$^1$, K.~Jim{\'e}nez-Garc{\'i}a$^{1,2}$ \& I.~B.~Spielman$^1$}

\begin{document}

\maketitle

\begin{affiliations}
 \item Joint Quantum Institute, National Institute of Standards and Technology, and University of Maryland, Gaithersburg, Maryland, 20899, USA
 \item Departamento de F\'{\i}sica, Centro de Investigaci\'{o}n y Estudios Avanzados del Instituto Polit\'{e}cnico Nacional, M\'{e}xico D.F., 07360, M\'{e}xico
\end{affiliations}

\begin{abstract}
Spin-orbit (SO) coupling -- the interaction between a quantum
particle's spin and its momentum -- is ubiquitous in nature, from
atoms to solids.  In condensed matter systems, SO coupling is
crucial for the spin-Hall effect\cite{Kato2004,Konig2007} and
topological insulators\cite{Kane2005,Bernevig2006,Hsieh2008}, which
are of extensive interest; it contributes to the electronic
properties of materials such as GaAs, and is important for
spintronic devices\cite{Koralek2009}. Ultracold atoms, quantum
many-body systems under precise experimental control, would seem to
be an ideal platform to study these fascinating SO coupled systems.
While an atom's intrinsic SO coupling affects its electronic
structure, it does not lead to coupling between the spin and the
center-of-mass motion of the atom. Here, we engineer SO coupling (with equal Rashba\cite{Bychkov1984} and
Dresselhaus\cite{Dresselhaus1955} strengths) in a neutral atomic
Bose-Einstein condensate by dressing two atomic spin states with a
pair of lasers\cite{Liu2009}. Not only is this the first SO coupling
realized in ultracold atomic gases, it is also the first ever
for bosons. Furthermore, in the presence of the laser
coupling, the interactions between the two dressed atomic spin
states are modified, driving a quantum phase transition from
a spatially spin-mixed state (lasers off) to a phase separated state
(above a critical laser intensity). The location of this transition
is in quantitative agreement with our theory.  This SO coupling --
equally applicable for bosons and fermions -- sets the stage to
realize topological insulators in fermionic neutral atom systems.
\end{abstract}

Quantum particles have an internal ``spin'' angular momentum;
this can be intrinsic for fundamental particles like electrons,
or a combination of intrinsic (from nucleons and electrons) and
orbital for composite particles like atoms. Spin-Orbit (SO) coupling
links a particle's spin to its motion, and generally appears for
particles moving in static electric fields, such as the nuclear
field of an atom or the crystal field in a material.  The coupling
results from the Zeeman interaction $-\vec{\mu}\cdot\vec{B}$ between
a particle's magnetic moment $\vec{\mu}$, parallel to the spin
$\vec{\sigma}$, and a magnetic field $\vec{B}$ present in the frame
moving with the particle. For example, Maxwell's equations dictate
that a static electric field $\vec{E}=E_0\hat z$ in the lab frame
(at rest) gives a magnetic field $\vec{B}_{\rm SO}=E_0(\hbar/m c^2)
(-k_y,k_x,0)$ in the frame of an object moving with momentum
$\hbar\vec{k}=\hbar(k_x,k_y,k_z)$, where $c$ is the speed of light
in vacuum and $m$ is the particle's mass. The resulting momentum
dependent Zeeman interaction $-\vec{\mu}\cdot\vec{B}_{\rm
SO}({\mathbf k})\propto \sigma_x k_y -\sigma_y k_x$ is known as the
Rashba\cite{Bychkov1984} SO coupling. In combination with the
Dresselhaus\cite{Dresselhaus1955} coupling $\propto -\sigma_x k_y -
\sigma_y k_x$, these describe two dimensional (2D) SO coupling in solids to first order.

In materials, the SO coupling strengths are generally intrinsic
properties, which are largely determined by the specific material
and the details of its growth, and thus only slightly adjustable in
the laboratory. 
We demonstrate SO coupling in a $\Rb87$
Bose-Einstein condensate (BEC) where a pair of Raman lasers create
a momentum-sensitive coupling between two internal atomic states.
This SO coupling is equivalent to that of an electronic system with
equal contributions of Rashba and Dresselhaus\cite{Liu2009}
couplings, and with a uniform magnetic field $\vec{B}$ in the $\hat
y$-$\hat z$ plane, which is described by the single particle
Hamiltonian
\begin{equation}\label{H_SO_extended}
\hat H\!=\!\frac{\hbar^2\hat{\mathbf k}^2}{2m}\check{1}
-\left[\vec{B}\!+\!\vec{B}_{\rm SO}(\hat {\mathbf k})\right]\cdot\vec\mu
 = \frac{\hbar^2\hat{\mathbf k}^2}{2m}\check{1}+\frac{\Omega}{2}\check{\sigma}_z+\frac{\delta}{2}\check{\sigma}_y+2 \alpha \hat{k}_x\check{\sigma}_y.
\end{equation}
$\alpha$ parametrizes the SO coupling strength; $\Omega=-g\mu_{\rm
B} B_z$ and $\delta=-g\mu_{\rm B} B_y$ result from the Zeeman fields
along $\hat z$ and $\hat y$, respectively; and
$\check\sigma_{x,y,z}$ are the 2$\times$2 Pauli matrices. Absent SO
coupling, electrons have group velocity $v_x=\hbar k_x/m$,
independent of their spin. With SO coupling, their velocity becomes
spin-dependent, $v_x=\hbar(k_x\pm 2\alpha m/\hbar^2)/m$ for spin
$\ket{\uparrow}$ and $\ket{\downarrow}$ electrons (quantized along
$\hat y$). In two recent experiments, this form of SO coupling was
engineered in GaAs heterostructures where confinement into 2D planes linearized GaAs's native cubic SO coupling to produce a Dresselhaus term, and asymmetries in the confining potential gave rise to Rashba coupling.  In one experiment a persistent
spin helix was found\cite{Koralek2009}, and in another the SO
coupling was only revealed by adding a Zeeman field\cite{Quay2010}. 

SO coupling for neutral atoms enables a range of exciting
experiments, and importantly, it is a key ingredient to
realize neutral atom topological insulators.  Topological insulators
are novel fermionic band insulators including integer-, and now
spin-quantum Hall states that insulate in the bulk, but conduct in
topologically protected quantized edge channels.  The first known
topological insulators -- integer quantum Hall
states\cite{Klitzing1980} -- require large magnetic fields that
explicitly break time-reversal symmetry.  In a seminal
paper\cite{Kane2005}, Kane and Mele showed that in some cases SO
coupling leads to zero magnetic field topological insulators
preserving time-reversal symmetry. Absent the bulk conductance that
plagues current materials, cold atoms can potentially realize these
insulators in their most pristine form, perhaps revealing their
quantized edge (in 2D) or surface (in 3D) states. To go beyond the
form of SO coupling we created, virtually any SO coupling, including
that needed for topological insulators, is possible with additional
lasers\cite{Ruseckas2005,Stanescu2007,Dalibard2010}.

\begin{figure}[t]
\begin{center}
\ifthenelse{\boolean{SubmittedVersion}}{}{\includegraphics[width=89 mm]{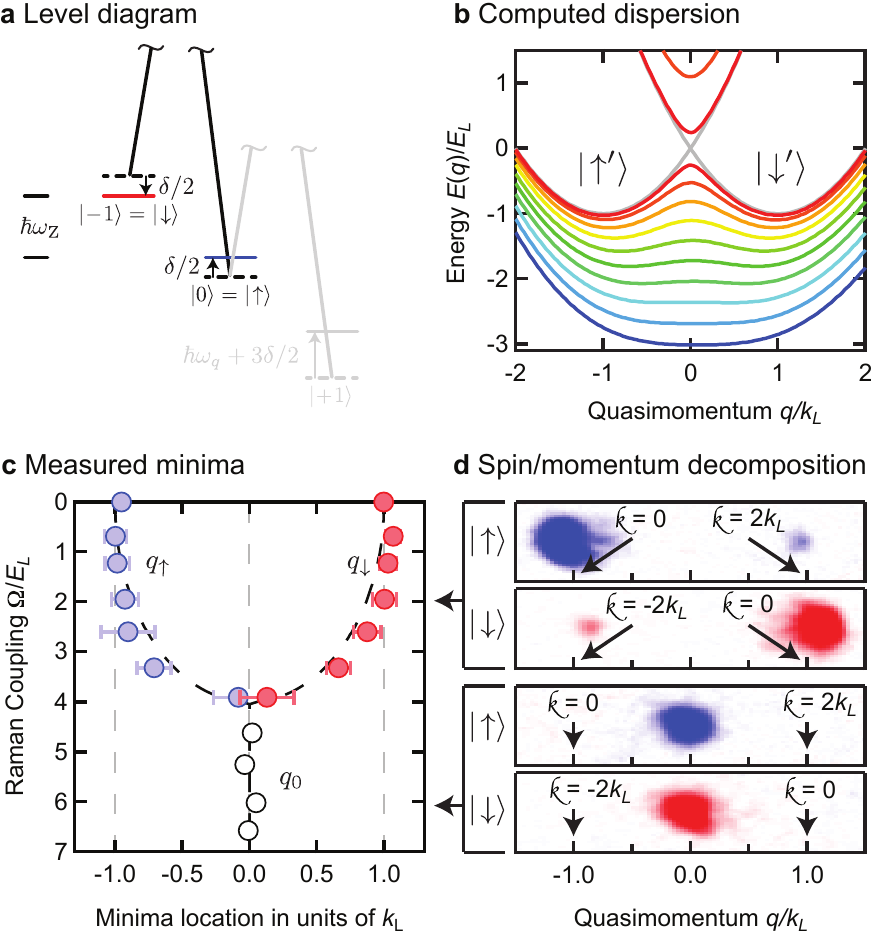}}
\end{center}
\caption{\textbf{Scheme for creating SO coupling}. \textbf{a}, Two
$\lambda=804.1\nm$ lasers (thick lines) coupled
states $\ket{F=1,m_F=0}=\ket{\uparrow}$ and $\ket{F=1,m_F=-1}=\ket{\downarrow}$, differing in energy by a
$\hbar\omega_Z$ Zeeman shift.  The lasers, with frequency
difference $\Delta\omega_L/2\pi = (\omega_Z+\delta/\hbar)/2\pi$,
were detuned $\delta$ from Raman resonance.
$\ket{m_F=0}$ and $\ket{m_F=+1}$ had a $\hbar(\omega_Z - \omega_q)$
energy difference; since $\hbar\omega_q=3.8 E_L$ is large, $\ket{m_F=+1}$ can be
neglected. \textbf{b}, Eigenenergies at $\delta=0$ for $\Omega=0$ (grey) to $5 E_L$.  When
$\Omega<4 E_L$ the two minima
correspond to \textit{dressed} spin states $\ket{\uparrow'}$ and
$\ket{\downarrow'}$. \textbf{c}, Quasimomentum
$q_{\uparrow,\downarrow}$ of $\ket{\uparrow',\downarrow'}$ versus $\Omega$ at $\delta=0$, corresponding to the
minima of $E_{-}(q)$.  Each point is averaged over about 10
experiments; the uncertainties are their standard deviation. \textbf{d}, Data for sudden laser turnoff: $\delta \approx
0$, $\Omega = 2 E_L$ (top image-pair), and $\Omega = 6 E_L$
(bottom pair). For $\Omega = 2 E_L$, $\ket{\uparrow'}$ consists of
$\ket{\uparrow,{\mathpzc k}_x\approx 0}$ and
$\ket{\downarrow,{\mathpzc k}_x\approx-2\kl}$, and
$\ket{\downarrow'}$ consists of $\ket{\uparrow,{\mathpzc k}_x\approx
2\kl}$ and $\ket{\downarrow,{\mathpzc k}_x\approx 0}$.}
\label{setup}
\end{figure}

To create SO coupling, we select two internal ``spin'' states from
within the $\Rb87$ $5{\rm S}_{1/2}$, $F=1$ ground electronic
manifold, and label them pseudo-spin up and down in analogy with an
electron's two spin states: $\ket{\uparrow}=\ket{F=1,m_F=0}$ and $\ket{\downarrow}=\ket{F=1,m_F=-1}$. A pair of $\lambda=804.1\nm$ Raman
lasers, intersecting at $\theta=90^\circ$ and detuned by $\delta$
from Raman resonance (Fig.~\ref{setup}a), couple these states with
strength $\Omega$; here $\hbar\kl=\sqrt{2}\pi\hbar/\lambda$ and
$E_L=\hbar^2 \kl^2/2m$ are the natural units of momentum and energy.  In this configuration, the atomic Hamiltonian is given by
Eq.~\ref{H_SO_extended}, with $k_x$ replaced by a quasimomentum $q$
and an overall $E_L$ energy offset.  $\Omega$ and
$\delta$ give rise to effective Zeeman fields along $\hat z$ and
$\hat y$, respectively. The SO coupling term $2E_L q
\check\sigma_{y}/k_L$ results from the laser geometry, and
$\alpha=E_L/k_L$ is set by $\lambda$ and $\theta$,
independent of $\Omega$ (see Methods).  In contrast with the electronic case, the
atomic Hamiltonian couples bare atomic states
$\ket{\uparrow,{\mathpzc k}_x=q+\kl}$ and $\ket{\downarrow,{\mathpzc
k}_x=q-\kl}$ with different velocities, $\hbar {\mathpzc
k}_x/m=\hbar(q\pm\kl)/m$.

\begin{figure}[t!]
\begin{center}
\ifthenelse{\boolean{SubmittedVersion}}{}{\includegraphics[width=89 mm]{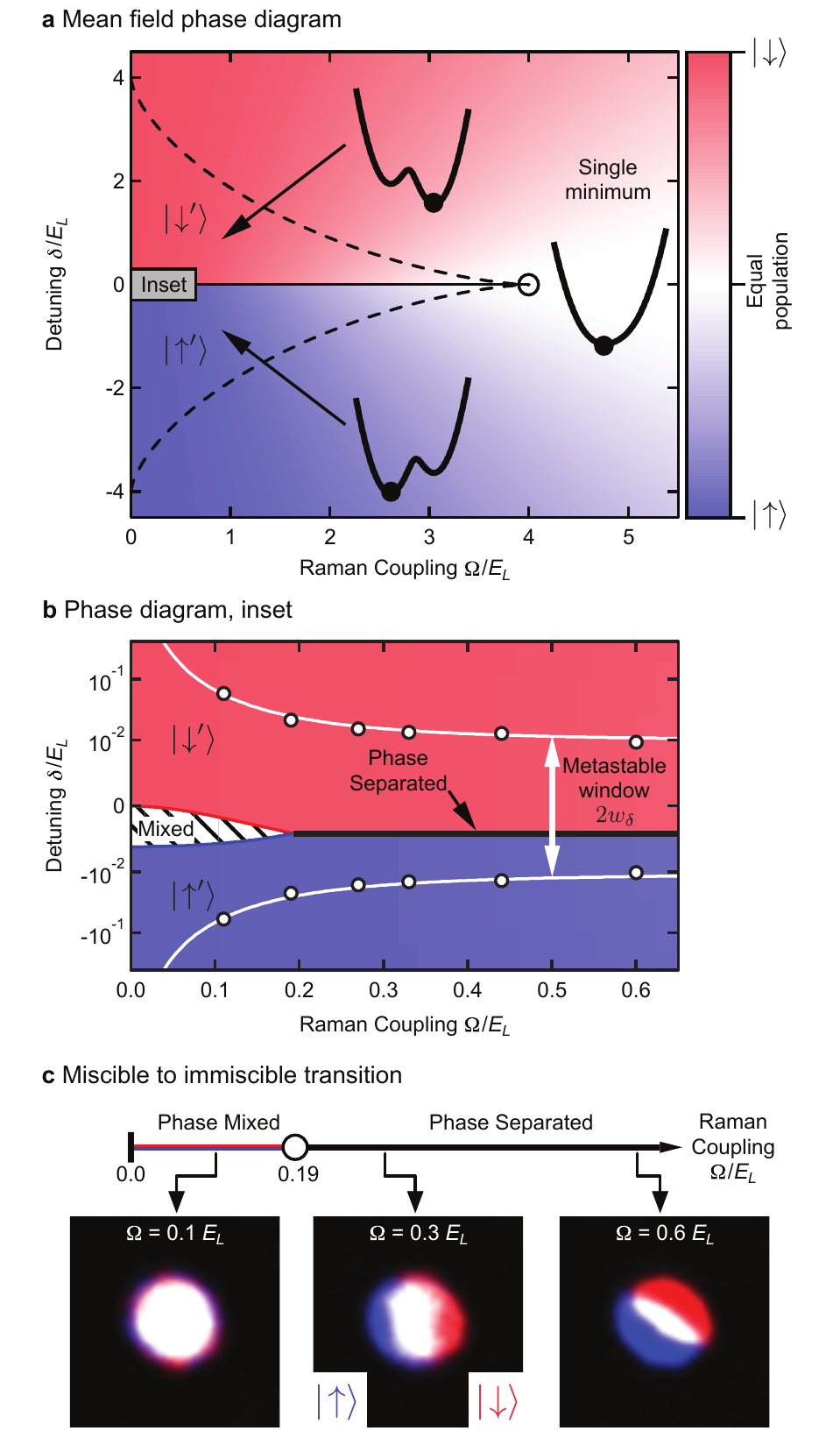}}
\end{center}
\caption{\textbf{Phases of a SO coupled BEC.} \textbf{a-b}, Mean
field phase diagrams for infinite homogeneous SO coupled $\Rb87$
BECs ($1.5\kHz$ chemical potential).  The background colors
indicate atom fraction in $\ket{\uparrow}$ and $\ket{\downarrow}$.  Between the dashed lines there are two dressed spin states,
$\ket{\uparrow'}$ and $\ket{\downarrow'}$.  \textbf{a}, Single particle phase diagram in the $\Omega-\delta$ plane. \textbf{b}, Phase diagram as modified by interactions.  The dots
represent a metastable region where the fraction of atoms
$f_{\uparrow',\downarrow'}$ remains largely unchanged for
$t_h=3\second$. \textbf{c}, Phase line for mixtures of
dressed spins and images after TOF (with populations $N_\uparrow\approx N_\downarrow$), mapped from
$\ket{\uparrow'}$ and $\ket{\downarrow'}$ showing the transition
from phase-mixed to phase-separated within the ``metastable window''
of detuning.} \label{PhaseDiagram}
\end{figure}

The spectrum, a new energy-quasimomentum dispersion of the SO
coupled Hamiltonian, is displayed in Fig.~\ref{setup}b at $\delta=0$
and for a range of couplings $\Omega$.  The dispersion is divided
into upper and lower branches $E_{\pm}(q)$, and we focus on
$E_-(q)$.  For $\Omega<4 E_L$ and small $\delta$ (see
Fig.~\ref{PhaseDiagram}a), $E_-(q)$ consists of a double-well in
quasi-momentum\cite{Higbie2004}, where the group velocity $\partial E_-(q)/\partial\hbar q$ is zero. States
near the two minima are dressed spin states, labeled as
$\ket{\uparrow'}$ and $\ket{\downarrow'}$. As $\Omega$ increases,
the two dressed spin states merge into a single minimum and the
simple picture of two dressed spins is inapplicable. Instead, that
strong coupling limit effectively describes spinless bosons with a
tunable dispersion relation\cite{Lin2009} with which we engineered
synthetic electric\cite{Lin2010} and magnetic fields\cite{Lin2009b}
for neutral atoms.

Absent Raman coupling, atoms with spins $\ket{\uparrow}$ and $\ket{\downarrow}$ 
spatially  mixed perfectly in a BEC. By increasing $\Omega$ we observed an
abrupt quantum phase transition to a new state where the two dressed
spins spatially separated, resulting from a modified effective
interaction between the dressed spins.

We studied SO coupling in oblate $\Rb87$ BECs with $\approx 1.8
\times 10^5$ atoms in a $\lambda=1064$~nm crossed dipole trap with
frequencies $(f_x,f_y,f_z)\approx(50,50,140)$~Hz. The bias magnetic
field $B_0 \hat y$ generated a $\omega_Z/2\pi\approx~4.81\MHz$
Zeeman shift between $\ket{\uparrow}$ and $\ket{\downarrow}$. The
Raman beams propagated along $\hat y \pm \hat x$ and had a constant
frequency difference $\Delta \omega_L/2\pi\approx 4.81$~MHz.  The
small detuning from Raman resonance $\delta=\hbar(\Delta
\omega_L-\omega_Z)$ was set by $B_0$, and  $\ket{m_F=+1}$ was
decoupled due to the quadratic Zeeman effect (see Methods).

\begin{figure}[t]
\begin{center}
\ifthenelse{\boolean{SubmittedVersion}}{}{\includegraphics[width= 89 mm]{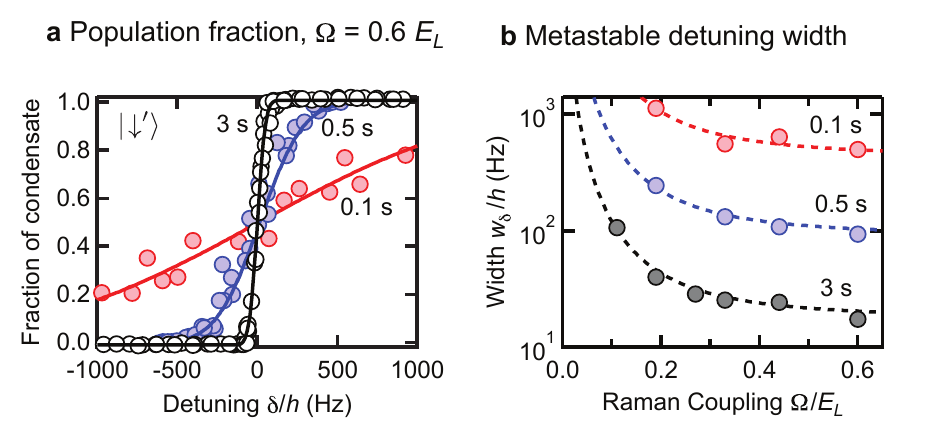}}
\end{center}
\caption{\textbf{Population relaxation.} \textbf{a}, Condensate
fraction $f_{\downarrow'}$ in $\ket{\downarrow'}$ versus detuning
$\delta$ at $t_h=0.1,0.5,\text{and } 3\second$ showing $w_\delta$
decrease with increasing $t_h$.  The solid curves are fits to the
error function from which we obtained the width $w_\delta$.
\textbf{b}, Width $w_\delta$ versus $\Omega$ at
$t_h=0.1,0.5,\text{and } 3\second$; the data fits well to $a\big[b +
(\Omega/E_L)^{-2}\big]$ (dashed curves).} \label{dynamics}
\end{figure}

We prepared BECs with an equal population of $\ket{\uparrow}$ and
$\ket{\downarrow}$ at $\Omega,\delta=0$, adiabatically
increased $\Omega$ to a final value up to $7E_L$ in $70\ms$, and
then allowed the system to equilibrate for $t_h=70\ms$. We abruptly
($t_{\rm off}< 1\us$) turned off the Raman lasers and the dipole
trap--thus projecting the dressed state onto their constituent bare
spin and momentum states--and absorption-imaged them after a
$30.1\ms$ time-of-flight (TOF).  For $\Omega>4 E_L$
(Fig.~\ref{setup}d), the BEC was located at the single minimum
$q_{\rm 0}$ of $E_-(q)$ with a single momentum component in each
spin state corresponding to the pair $\left\{\ket{\uparrow,q_{\rm 0}+\kl},\ket{\downarrow,q_{\rm
0}-\kl}\right\}$.  However, for
$\Omega<4 E_L$ we observed two momentum components in each spin
state, corresponding to the two minima of $E_-(q)$ at
$q_{\uparrow}$ and $q_{\downarrow}$.  The agreement between the data
(symbols), and the expected minima-locations (curves), demonstrates
the existence of the SO coupling associated with the Raman dressing.
We maintained $\delta\approx 0$ when turning on $\Omega$ by making
equal populations in bare spins $\ket{\uparrow},\ket{\downarrow}$
(see Fig.~\ref{setup}d).


We experimentally studied the low temperature phases of these
interacting SO coupled bosons as a function of $\Omega$ and
$\delta$.  The zero-temperature mean-field phase diagram (Fig.~\ref{PhaseDiagram}a,b) includes phases composed of: a single
dressed spin state, a spatial mixture of both dressed spin states,
and coexisting but spatially phase-separated dressed spins.

This phase diagram can be largely understood from non-interacting
bosons condensing into the lowest energy single particle state, and
can be divided into three regimes (Fig.~\ref{PhaseDiagram}a). In the
region of positive detuning marked $\ket{\downarrow'}$, there are
double minima at $q=q_{\uparrow},q_{\downarrow}$ in $E_{-}(q)$ with
$E_{-}(q_{\downarrow})<E_{-}(q_{\uparrow})$ and the bosons condense
at $q_{\downarrow}$. In the region marked $\ket{\uparrow'}$ the
reverse holds. The energy difference between the two minima is
$\Delta(\Omega,\delta)=E_-(q_{\uparrow})-E_-(q_{\downarrow})\approx
\delta$ for small $\delta$ (see Methods). In the third ``single
minimum'' regime, the atoms condense at the single minimum $q_0$.
These dressed spins act as free particles with group velocity $\hbar
K_x/m$ (with an effective mass $m^*\approx m$, for
small $\Omega$), where $K_x=q-q_{\uparrow,\downarrow,0}$ for the
different minima.

We investigated the phase diagram using BECs with initially equal
spin populations prepared as described previously, but with
$\delta\neq 0$ and $t_h$ up to $3$~s.  We probed the atoms after
abruptly removing the dipole trap, and then ramping
$\Omega\rightarrow0$ in $1.5\ms$. This approximately mapped
$\ket{\uparrow'}$ and $\ket{\downarrow'}$ back to their undressed
counterparts
$\ket{\uparrow}$
and
$\ket{\downarrow}$
(see Methods). We absorption-imaged the atoms
after a $30\ms$ TOF, during the last $20\ms$ of which a
Stern-Gerlach magnetic field gradient along $\hat y$ separated the
spin components.

Figure~\ref{dynamics}a shows the condensate fraction
$f_{\downarrow'} =N_{\downarrow'} / (N_{\downarrow'} +
N_{\uparrow'})$ in $\ket{\downarrow'}$ at $\Omega=0.6E_L$ as a
function of $\delta$, at $t_h=0.1$, $1$ and $3$~s, where
$N_{\uparrow'}$ and $N_{\downarrow'}$ denote the number of condensed atoms in $\ket{\uparrow'}$ and $\ket{\downarrow'}$, respectively. The BEC
is all $\ket{\uparrow'}$ for $\delta\lesssim0$ and all
$\ket{\downarrow'}$ for $\delta\gtrsim0$, but both dressed spin populations
substantially coexisted for detunings within $\pm w_{\delta}$ (obtained by
fitting $f_{\downarrow'}$ to the error function where
$\delta=\pm w_{\delta}$ corresponds to $f_{\downarrow'}=0.50 \pm
0.16$). Figure~\ref{dynamics}b shows $w_{\delta}$ versus $\Omega$
for hold times $t_h$. $w_{\delta}$ decreases with $t_h$; even by our
longest $t_h=3$~s it has not reached equilibrium.  


Conventional $F=1$ spinor BECs have been studied in $\Na23$ and
$\Rb87$ without Raman coupling\cite{Stenger1998,Chang2004,Ho1998}. For our 
$\ket{\uparrow}$ and $\ket{\downarrow}$ states, the interaction energy
depends on the local density in each spin state, and is described by
\begin{align*}
\hat H_{\rm I}\! =& \frac{1}{2}\!\int\!d^3r \bigg[\!
\left(c_0\!+\!\frac{c_2}{2}\right)\!\left(\hat \rho_{\uparrow}\!+\!\hat \rho_{\downarrow} \right)^2
+ \frac{c_2}{2}\!\left(\hat \rho^2_{\downarrow}\!-\!\hat
\rho^2_{\uparrow} \right) + (c_2\!+\!c^\prime_{\uparrow \downarrow})
\hat \rho_{\uparrow} \hat \rho_{\downarrow} \bigg],
\end{align*}
where $\hat \rho_{\uparrow}$ and $\hat \rho_{\downarrow}$ are
density operators for $\ket{\uparrow}$ and $\ket{\downarrow}$. In
the $\Rb87$ $F\!=\!1$ manifold, the spin independent interaction is
$c_0 = 7.79\times10^{-12}\Hz\cm^3$, the spin dependent interaction\cite{Widera2006}
is $c_2 = -3.61\times10^{-14}\Hz\cm^3$, and $c^\prime_{\uparrow \downarrow}=0$.  Since $|c_0|\gg |c_2|$ the interaction is almost
spin independent, but because $c_2<0$, the two-component mixture of
$\ket{\uparrow}$ and $\ket{\downarrow}$ has a spatially mixed ground
state (is miscible). When $\hat H_{\rm I}$ is re-expressed in terms
of the dressed spin states, $c^\prime_{\uparrow \downarrow}\approx c_0
\Omega^2/(8E_L^2)$ is nonzero and corresponds to an effective
interaction between $\ket{\uparrow'}$ and $\ket{\downarrow'}$. This
modifies the ground state of our SO coupled BEC (mixtures of
$\ket{\uparrow'}$ and $\ket{\downarrow'}$) from phase-mixed to
phase-separated above a critical Raman coupling strength $\Omega_c$.
This transition lies outside the common single-mode
approximation\cite{Chang2004}.

The effective interaction between $\ket{\uparrow'}$ and
$\ket{\downarrow'}$ is an exchange energy resulting from the
non-orthogonal spin part of $\ket{\uparrow'}$ and
$\ket{\downarrow'}$ (see Methods): a spatial mixture produces total
density modulations\cite{Higbie2004} with wavevector $2\kl$ in analogy with the
spin-textures of the electronic case\cite{Koralek2009}. These
increase the state-independent interaction energy in $\hat
H_{\rm I}$ wherever the two dressed spins spatially overlap,
contributing to the $c^\prime_{\uparrow \downarrow}$ term. (Such a
term does not appear for rf-dressed states, which are always
spin-orthogonal.)
Because $c^\prime_{\uparrow\downarrow}$ and $c_2$ have opposite
sign here, the dressed BEC can go from miscible to immiscible, at the miscibility threshold\cite{Stenger1998} for a two-component
BEC $c_0 + c_2 + c^\prime_{\uparrow\downarrow}/2 = \sqrt{c_0
(c_0+c_2)}$, when $\Omega=\Omega_c$ (this result is in agreement with an independent theory presented in Ref.~\cite{Ho2010}).

Figure~\ref{PhaseDiagram}b depicts the mean field phase diagram
\textit{including} interactions, computed by minimizing the
interaction energy $H_{\rm I}$ plus the single particle detuning
$\Delta(\Omega,\delta)\approx\delta$.  This phase diagram adds to the non-interacting picture both mixed (hashed) and phase-separated (bold line) regimes.  The $c_2\!\left(\hat \rho^2_{\downarrow}\!-\!\hat
\rho^2_{\uparrow} \right)/2$ term in $\hat
H_{\rm I}$ implies that the energy difference between a $\ket{\uparrow}$ BEC and a $\ket{\downarrow}$ BEC is proportional $N^2 c_2$.  The detuning required to compensate for this difference slightly displaces the symmetry point of the phase diagram downwards.  As evidenced by the width of the metastable window $2w_{\delta}$ in
Fig.~\ref{PhaseDiagram}b, for $|\delta|<w_{\delta}$ the
spin-population does not have time to relax to equilibrium.
Since the miscibility condition does not depend on atom
number, the phase line in Fig.~\ref{PhaseDiagram}c shows the
system's phases for $\left|\delta\right|<w_\delta$: phase-mixed for $\Omega<\Omega_c$ and
phase-separated for $\Omega>\Omega_c$ where
$\Omega_c\approx\sqrt{-8c_2/c_0}E_L\approx0.19 E_L$.

\begin{figure}[t]
\begin{center}
\ifthenelse{\boolean{SubmittedVersion}}{}{\includegraphics[width= 89 mm]{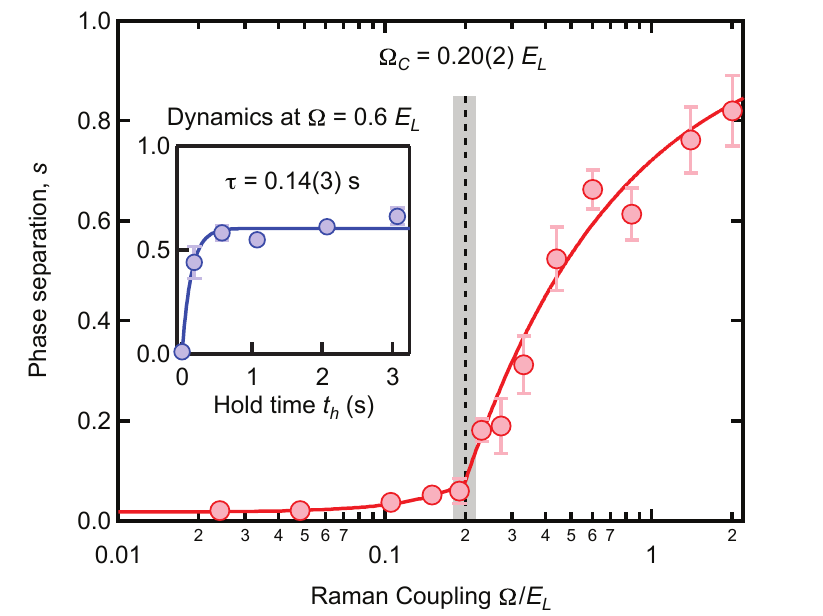}}
\end{center}
\caption{\textbf{Miscible to immiscible phase transition.} Phase separation $s$ versus $\Omega$ with
$t_h=3$~s; the solid curve is a fit to the function described in the
text.  The power law component of the fit has an exponent $a=0.75\pm0.07$; this is not a critical exponent, but instead results from the decreasing size of the domain wall between the regions of $\ket{\uparrow^\prime}$ and $\ket{\downarrow^\prime}$ as $\Omega$ increases.  Each point represents an average over 15 to 50 realizations and the uncertainties are the standard deviation. \textbf{Inset}, Phase separation $s$ versus $t_h$ with
$\Omega=0.6 E_L$ fit to an exponential showing the rapid
$0.14(3)\second$ time scale for phase separation. } \label{transition}
\end{figure}

We measured the miscibility of the dressed spin components from
their spatial profiles after TOF, for $\Omega=0$ to $2 E_L$ and
$\delta\approx 0$ such that $N_{\rm T\uparrow'}\approx N_{\rm
T\downarrow'}$, where $N_{\rm T\uparrow',\downarrow'}$ is the total
atom number including both the condensed and thermal components in
$\ket{\uparrow'},\ket{\downarrow'}$. For each TOF image, we
numerically recentered the Stern-Gerlach-separated spin
distributions (Fig.~\ref{PhaseDiagram}c, and see Methods), giving
condensate densities $n_{\uparrow'}(x,y)$ and $n_{\downarrow'}(x,y)$.  Since the
self-similar expansion of BECs released from harmonic traps
essentially magnifies the in-situ spatial spin distribution, these
reflect the in-situ densities\cite{Hall1998}.

A dimensionless metric $s=1-\langle n_{\uparrow'}
n_{\downarrow'}\rangle/\left(\langle n_{\uparrow'}^2\rangle\langle
n_{\downarrow'}^2\rangle\right)^{1/2}$ quantifies the degree of phase separation
($\langle\ldots \rangle$ is the spatial average over a single image):
$s=0$ for any perfect mixture $n_{\uparrow'}(x,y) \propto
n_{\downarrow'}(x,y)$, and $s=1$ for complete phase separation.
Figure~\ref{transition} displays $s$ versus Raman coupling $\Omega$
with a hold time $t_h=3$~s, showing that $s\approx 0$ for small
$\Omega$ (as expected given our miscible bare spins) and $s$
abruptly increases above a critical $\Omega_c$.
The inset to Fig.~\ref{transition} plots $s$ as a function of time, showing that
$s$ reaches steady-state in $0.14(3)\second\ll t_h$.  To obtain
$\Omega_c$, we fit the data in Fig.~\ref{transition} to a slowly
increasing function below $\Omega_c$ and the power-law $1-(\Omega/\Omega_c)^{-a}$ above $\Omega_c$.  The
resulting $\Omega_c=0.20(2) E_L $ is in agreement with the mean
field prediction $\Omega_c=0.19 E_L$. This demonstrates a quantum
phase transition for a two-component SO coupled BEC, from miscible
when $\Omega<\Omega_c$ to immiscible when $\Omega > \Omega_c$.

Even below $\Omega_c$, $s$ slowly increased with increasing $\Omega$.  To understand this effect, we numerically solved the 2D spinor Gross-Pitaevskii equation in the presence of a trapping potential.  This demonstrated that the differential interaction term $c_2\!\left(\hat \rho^2_{\downarrow}\!-\!\hat \rho^2_{\uparrow} \right)/2$ in $\hat H_I$ favors slightly different density profiles for each spin component, while the $(c_2+c^\prime_{\uparrow\downarrow}) \hat\rho_{\uparrow} \hat\rho_{\downarrow}$ term favors matched profiles.  Thus, as $c_2+c^\prime_{\uparrow\downarrow}$ approached zero from below this balancing effect decreased, leading $s$ to increase.

An infinite system should fully phase separate ($s=1$) for all $\Omega>\Omega_c$.  In our finite system, the boundary between the phase separated spins, set by the spin-healing length ($\xi_s=\sqrt{\hbar^2/2m\left|c_2+c^\prime_{\uparrow\downarrow}\right|n}$, where $n$ is the local density), can be comparable to the system size.  We interpret the increase of $s$ above $\Omega_c$ as resulting from the decrease of $\xi_s$ with increasing $\Omega$. 

We realized SO coupling in a $\Rb87$ BEC, and observed a quantum
phase transition from spatially mixed to spatially separated. By
operating at lower magnetic field (with a smaller quadratic Zeeman
shift), our method extends to the full $F=1$ or $F=2$ manifold of
$\Rb87$ or $\Na23$, enabling a new kind of tuning for spinor
BECs, without the losses associated with Feshbach
tuning\cite{Erhard2004}.  Such modifications may allow access to the
expected non-abelian vortices in some $F=2$
condensates\cite{Kobayashi2009}. Since our SO coupling is in the
small $\Omega$ limit, this technique is practical for fermionic
$^{40}{\rm K}$, with its smaller fine-structure splitting and thus
larger spontaneous emission rate\cite{Goldman2010a}. When the Fermi
energy lies in the gap between the lower and upper bands (e.g.,
Fig.~\ref{setup}b) there will be a single Fermi surface; this
situation can induce $p$-wave coupling between
fermions\cite{Zhang2008} and more recent work anticipates the
appearance of Majorana fermions\cite{Sau2010}.

\begin{methodssummary}

\subsection{System preparation}
Our experiments began with nearly pure $\approx 1.8\times 10^5$
atom $\Rb87$ BECs in the $\ket{F=1,m_F=-1}$ state\cite{Lin2009a}
confined in a crossed optical dipole trap.  The trap consisted of a
pair of $1064\nm$ laser beams propagating along $\hat{x}-\hat{y}$
($1/e^2$ radii of $w_{\hat{x}+\hat{y}}\approx 120\micron$ and
$w_{\hat z}\approx 50\micron$) and $-\hat{x}-\hat{y}$ ($1/e^2$ radii
of $w_{\hat{x}-\hat{y}}\approx w_{\hat z}\approx 65\micron$).

We prepared equal mixtures of $\ket{F=1,m_F=-1}$ and $\ket{1,0}$
using an initially off resonant rf magnetic field $B_{\text{rf}}(t)
\hat{x}$.  We adiabatically ramped $\delta$ to $\delta \approx0$ in
$15\ms$, decreased the rf coupling strength $\Omega_{\text{rf}}$ to
about $150\Hz \ll \hbar\omega_q$ in $6\ms$, and suddenly turned off
$\Omega_{\text{rf}}$, projecting the BEC into an equal
superposition of $\ket{m_F=-1}$ and $\ket{m_F=0}$. We subsequently
ramped $\delta$ to its desired value in $6\ms$ and then linearly
increased the intensity of the Raman lasers from zero to
the final coupling $\Omega$ in $70\ms$.

\subsection{Magnetic fields}
Three pairs of Helmholtz coils, orthogonally aligned along
$\hat{x}+\hat{y}$, $\hat{x}-\hat{y}$ and $\hat{z}$, provided bias
fields $(B_{x+y},B_{x-y},\text{and } B_{z})$.  By monitoring the
$\ket{F=1,m_F=-1}$ and $\ket{1,0}$ populations in a nominally
resonant rf dressed state, prepared as above, we observed a
short-time (below $\approx10$ minutes) RMS field stability $g
\mu_{\rm B} B_{\rm RMS}/h\lesssim 80\Hz$.  The field drifted slowly
on longer time scales (but changed abruptly when unwary colleagues
entered through our laboratory's ferromagnetic doors).  We compensated for
the drift by tracking the rf and Raman resonance conditions.

Due to the small energy scales involved in the experiment, it was
crucial to minimize magnetic field gradients.  We detected stray
gradients by monitoring the spatial distribution of
$\ket{m_F=-1}$-$\ket{m_F=0}$ spin mixtures after TOF.  Small
magnetic field gradients caused this otherwise miscible mixture
to phase separate along the direction of the gradient.  We canceled
the gradients in the $\hat x\!-\!\hat y$ plane with two pairs of
anti-Helmholtz coils, aligned along $\hat x\!+\!\hat y$ and $\hat x\!-\!\hat
y$, to $g\mu_B B'/h\lesssim 0.7\Hz/\mu\text{m}$.
\end{methodssummary}

\begin{addendum}
 \item We thank E.~Demler, T.-L.~Ho, and H.~Zhai for conceptual input; and we appreciate conversations with J.~V.~Porto, and W.~D.~Phillips.  This work was partially supported by ONR, ARO with funds from the DARPA OLE program, and the NSF through the Physics Frontier Center at JQI.  K.J.-G. acknowledges CONACYT.
 \item[Competing Interests] The authors declare that they have no
competing financial interests.
 \item[Author Contributions] All authors contributed to writing of the manuscript.  Y.-J.~L. lead the data taking effort in which K.~J.-G. participated.  I.~B.~S. conceived the experiment; performed numerical and analytic calculations; and supervised this work.
 \item[Author Information] Correspondence and requests for materials
should be addressed to\\ I.~B.~S.~(ian.spielman@nist.gov).
\end{addendum}

\ifthenelse{\boolean{SubmittedVersion}}{\processdelayedfloats}{\cleardoublepage}

\begin{methods}

\subsection{SO coupled Hamiltonian}
Our system consisted of a $F=1$ BEC with a bias magnetic field along
$\hat y$ at the intersection of two Raman laser beams propagating
along $\hat{x}+\hat{y}$ and $-\hat{x}+\hat{y}$ with angular
frequencies $\omega_L$ and $\omega_L+\Delta\omega_L$, respectively.  The
rank-1 tensor light shift of these beams produced an effective
Zeeman magnetic field along the $z$ direction with Hamiltonian
$\hat H_R=\Omega_R\check\sigma_{3,z} \cos(2\kl \hat x + \Delta\omega_L t)$,
where $\check\sigma_{3,x,y,z}$ are the $3\times3$ Pauli matrices and we
define $\check 1_3$ as the $3\times3$ identity matrix.  If we take
$\hat y$ as the natural quantization axis (by expressing the Pauli
matrices in a rotated basis $\check\sigma_{3,y}\rightarrow\check\sigma_{3,z}$, $\check\sigma_{3,x}\rightarrow\check\sigma_{3,y}$, and $\check\sigma_{3,z}\rightarrow\check\sigma_{3,x}$) and make the
rotating wave approximation, the Hamiltonian for spin states
$\left\{\ket{m_F=+1},\ket{0},\ket{-1} \right\}$ in the frame rotating at $\Delta \omega_L$ is
\begin{eqnarray}
\hat H_3 & = & \frac{\hbar^2\hat{\mathbf{k}}^2}{2m}\check 1_3 +
\left(
\begin{array}{ccc}
3\delta/2+\hbar\omega_q & 0 & 0 \\0 & \delta/2 & 0 \\0 & 0 &
-\delta/2
\end{array}
\right) + \\
& & \frac{\Omega_R}{2}\check\sigma_{3,x}\cos(2 \kl \hat x) - \frac{\Omega_R}{2}\check\sigma_{3,y}\sin(2 \kl \hat x).\nonumber
\end{eqnarray}
As we justify below, $\ket{m_F=+1}$ can be neglected for large
enough $\hbar\omega_q$, which gives the effective two-level
Hamiltonian
\begin{eqnarray}
\hat H_2 & = & \frac{\hbar^2\hat{\mathbf{k}}^2}{2m}\check 1 +
\frac{\delta}{2}\check\sigma_z +
\frac{\Omega}{2}\check\sigma_{x}\cos(2 \kl \hat x) -
\frac{\Omega}{2}\check\sigma_{y}\sin(2 \kl \hat x)\nonumber
\end{eqnarray}
for the pseudo-spin $\ket{\uparrow}=\ket{m_F=0}$ and
$\ket{\downarrow}=\ket{-1}$ where $\Omega = \Omega_R/\sqrt{2}$.  After a \textit{local}
pseudo-spin rotation by $\theta(\hat x) = 2 \kl \hat x$ about the
pseudo-spin $\hat z$ axis followed by a global pseudo-spin rotation
$\check\sigma_z\rightarrow\check\sigma_y$, $\check\sigma_y\rightarrow\check\sigma_x$, and $\check\sigma_x\rightarrow\check\sigma_z$, the $2\times2$ Hamiltonian takes the SO coupled
form
\begin{equation}\label{H_SO_extended2}
\hat H_{2} = \frac{\hbar^2\hat{\mathbf k}^2}{2m}\check{1}+\frac{\Omega}{2}\check{\sigma}_z+\frac{\delta}{2}\check{\sigma}_y+2 \frac{\hbar^2\kl\hat{k}_x}{2m}\check{\sigma}_y + E_L\check{1}.\nonumber
\end{equation}
The SO term linear in $\hat k_x$ results from the non-commutation of
the spatially-dependent rotation about the pseudo-spin $z$ axis and the
kinetic energy.

\subsection{Effective two-level system}
For atoms in $\ket{m_F=-1}$ and $\ket{m_F=0}$ with velocities $\hbar
{\mathpzc k}_x/m\approx0$ and Raman-coupled near resonance, $\delta
\approx0$, the $\ket{m_F=+1}$ state is detuned from resonance owing to the
$\hbar \omega_q=3.8E_L$ quadratic Zeeman shift. For
$\delta/4E_L\ll1$ and $\Omega <4E_L$, $\Delta(\Omega,\delta)\approx
\delta[1-(\Omega/4E_L)^2]^{1/2}$.

\subsection{Effect of the neglected state}
In our experiment, we focused on the two level system formed by the
$\ket{m_F=-1}$ and $\ket{m_F=0}$ states. We verified the validity of
this assumption by adiabatically eliminating the $\ket{m_F=+1}$
state from the full three level problem.  To second order in
$\Omega$, this procedure modifies the detuning $\delta$ and SO
coupling strength $\alpha$ in Eq.~\ref{H_SO_extended} by
\begin{eqnarray}
\delta^{(2)}&=&\left(\frac{\Omega}{2}\right)^2\frac{1}{4 E_L+\hbar\omega_q}\approx\frac{1}{32}\frac{\Omega^2}{E_L}\nonumber \\
\alpha^{(2)}&=&\left(\frac{\Omega}{2}\right)^2\frac{\alpha}{(4E_L+\hbar\omega_q)^2}\approx\frac{\alpha}{256}\left(\frac{\Omega}{E_L}\right)^2.\nonumber
\end{eqnarray}
In these expressions, we have retained only largest term in a
$1/\omega_q$ expansion.  In our experiment, where $\hbar\omega_q=3.8E_L$, $\delta$ is substantially changed at our largest coupling $\Omega=7 E_L$.  To maintain the desired detuning $\delta$ in the simple 2-level model (i.e., $\Delta\approx \delta+\delta^{(2)}=0$ in Fig.~\ref{setup}c), we changed $g\mu_{\rm B} B_0$ by as much as $3E_L$ to compensate for $\delta^{(2)}$.  We
did not correct for the always small change to $\alpha$.

Although both terms are small at the $\Omega=0.2 E_L$ transition
from miscible to immiscible, slow drifts in $B_0$ prompted us to
locate $\Delta=0$ empirically from the equal population condition,
$N_{\rm T\uparrow'}=N_{\rm T\downarrow'}$.  As a result, $\delta$ in
Eq.~\ref{H_SO_extended} implicitly includes the perturbative
correction $\delta^{(2)}$.

\subsection{Origin of the effective interaction term}
The additional $c^\prime_{\uparrow\downarrow}$ term in the
interaction Hamiltonian for dressed spins directly results from
transforming into the basis of dressed spins, which are
\begin{align}
\ket{\uparrow',K_x} \approx& \ket{\uparrow,{\mathpzc k}_x=K_x+q_{\uparrow}+\kl}-\epsilon\ket{\downarrow,{\mathpzc k}_x=K_x+q_{\uparrow}-\kl},\ {\rm and} \nonumber\\
\ket{\downarrow',K_x}\approx&\ket{\downarrow,{\mathpzc
k}_x=K_x+q_{\downarrow}-\kl}-\epsilon\ket{\uparrow,{\mathpzc
k}_x=K_x+q_{\downarrow}+\kl},\label{DressedSpins}
\end{align}
where $\hbar K_x/m$ is the group velocity, $K_x=q-q_{\uparrow}$ for
$\ket{\uparrow'}$ and $K_x=q-q_{\downarrow}$ for
$\ket{\downarrow'}$, and $\epsilon = \Omega/8E_L\ll1$.  Thus, in
second quantized notation, the dressed field operators transform
according to
\begin{align*}
\hat \psi_\uparrow(r) &= \hat \psi_{\uparrow'}(r) + \epsilon
e^{2i\kl x}\hat \psi_{\downarrow'}(r)
\end{align*}
and
\begin{align*}
\hat \psi_\downarrow(r) &= \hat \psi_{\downarrow'}(r) + \epsilon
e^{-2i\kl x}\hat \psi_{\uparrow'}(r),
\end{align*}
where $q_\uparrow\approx-\sqrt{1-4\epsilon ^2}\kl\approx-\kl$ and
$q_\downarrow\approx\sqrt{1-4\epsilon ^2}\kl\approx\kl$. Inserting
the transformed operators into
\begin{eqnarray}
\hat H_{\rm I} = \frac{1}{2}\int d^3r \bigg[ \left(c_0 +
\frac{c_2}{2}\right)\left(\hat \rho_{\downarrow} + \hat
\rho_{\uparrow} \right)^2 + \frac{c_2}{2} \left(\hat
\rho^2_{\downarrow} - \hat \rho^2_{\uparrow} \right) + c_2 \hat
\rho_{\downarrow} \hat \rho_{\uparrow} \bigg]\nonumber
\end{eqnarray}
gives the interaction Hamiltonian for dressed spins which can be
understood order-by-order (both $c_2/c_0$ and $\epsilon$ are treated
as small parameters).  In this analysis, the terms proportional to
$c_2$ are unchanged to order $c_2/c_0$, and we only need to evaluate
the transformation of the spin-independent term (proportional to
$c_0$). At $O(\epsilon)$ and $O(\epsilon^3)$ all the terms in the
expansion include high spatial frequency $e^{\pm2i\kl x}$
or $e^{\pm4i\kl x}$ prefactors. For density distributions
that vary slowly on the $\lambda/2$ length scale these average to
zero. The $O(\epsilon^2)$ term, however, has terms without these
modulations, and is
\begin{align*}
\hat H_{\rm I}^{(\epsilon^2)} =& \frac{1}{2}\int d^3r\left( 8c_0\epsilon^2
\hat\psi^\dagger_{\downarrow'}\hat\psi^{\dagger}_{\uparrow'}
\hat\psi_{\downarrow'} \hat\psi_{\uparrow'} \right),
\end{align*}
giving rise to $c^\prime_{\uparrow\downarrow} =
c_0\Omega^2/(8E_L^2)$.

\subsection{Mean field phase diagram} We compute the mean-field phase
diagram for a ground state BEC composed of a mixture of dressed
spins in an infinite homogeneous system. This applies to our atoms in
a harmonic trap in the limit of $R\gg\xi_s$, where $R$ is the system
size, $\xi_s=\sqrt{\hbar^2/2m|c_2+c^\prime_{\uparrow\downarrow}|n}$ is the spin healing length and
$n$ is the density. We first minimize the interaction energy
$\hat H_{\rm I}$ at fixed $N_{\uparrow',\downarrow'}$, with an effective interaction
$c^{\prime}_{\uparrow \downarrow}$ as a function
of $\Omega$. The two dressed spins are either
phase-mixed, both fully occupying the system's volume $V$, or
phase-separated with a fixed total volume constraint
$V=V_{\uparrow'} + V_{\downarrow'}$. For the phase-separated case,
minimizing the free energy gives the volumes $V_{\uparrow'}$ and
$V_{\downarrow'}$, determined by $N_{\uparrow',\downarrow'}$ and
$V$. The interaction energy of a phase-mixed state is smaller than
that of a phase-separated state for the miscibility condition $c_0 +
c_2 + c^\prime_{\uparrow\downarrow}/2 < \sqrt{c_0 (c_0+c_2)}$,
corresponding to $\Omega<\Omega_c$. This condition is independent of
$N_{\uparrow',\downarrow'}$: for any $N_{\uparrow',\downarrow'}$ the
system is miscible at $\Omega<\Omega_c$. Then, at a given $\Omega$,
we minimize the sum of the interaction energy and the
single-particle energy from the Raman detuning,
$(N_{\uparrow'}-N_{\downarrow'})\delta/2$, allowing
$N_{\uparrow',\downarrow'}$ to vary. For the miscible case 
$(\Omega<\Omega_c)$, the BEC is a mixture with fraction $N_{\downarrow'}/(N_{\uparrow'}+N_{\downarrow'})\in(0,1)$ only in the range of detuning $\delta\in(\delta_0-W_{\delta},\delta_0+W_{\delta})$, where
$\delta_0=c_2n/2$, $W_{\delta}=|\delta_0|(1-\Omega/\Omega_c)^{1/2}$
and $n=(N_{\uparrow'}+N_{\downarrow'})/V$. For the immiscible case
($\Omega>\Omega_c$), $W_{\delta}=(c_2/8c_0)c_2n$ is negligibly small
compared to $c_2n$.

Figure~\ref{PhaseDiagram}b shows the mean field phase diagram as a
function of $(\Omega,\delta)$, where $\delta/E_L$ is displayed with
a quasi-logarithmic scaling, ${\rm sgn}(\delta/E_L)\left[\log_{10}\left(|\delta/E_L|+|\delta_{\rm min}/E_L|\right)-\log_{10}|\delta_{\rm{
min}}/E_L|\right]$, in order to display $\delta$ within the range of
interest. This scaling function smoothly evolves from logarithmic
for $|\delta|\gg \delta_{\rm{min}}$, $\approx {\rm
sgn}(\delta/E_L)\log_{10}|\delta/E_L|$, to linear for $|\delta|\ll
\delta_{\rm min}$, $\approx \delta$, where $\delta_{\rm
min}/E_L=0.001 E_L=1.5~$Hz.

In our measurement of the dressed spin fraction $f_{\downarrow'}$
(see Fig.~\ref{dynamics}a), $\delta=0$ is determined from the $N_{\rm
T\uparrow'}=N_{\rm T\downarrow'}$ condition. We identify this
condition as $\delta=\delta_0$ and apply it for all hold time $t_h$.
Because $|\delta_0|\approx 3~$Hz is below our $\approx 80~$Hz RMS
field noise, we are unable to distinguish $\delta_0$ from 0.

\subsection{Recombining TOF images of dressed spins} To probe the
dressed spin states (Eq.~\ref{DressedSpins}), each of which is a spin
and momentum superposition, we adiabatically mapped them into bare
spins, $\ket{\uparrow,{\mathpzc k}_x=q_{\uparrow}+\kl}$ and
$\ket{\downarrow,{\mathpzc k}_x=q_{\downarrow}-\kl}$, respectively.
Then, in each image outside a $\approx90\micron$ radius disk
containing the condensate for each spin distribution, we fit $n_{\rm
T{\uparrow'},\rm T{\downarrow'}}(x,y)$ to a gaussian modeling the
thermal background and subtracted that fit from $n_{\rm
T{\uparrow'},\rm T{\downarrow'}}(x,y)$ to obtain the condensate 2D
density $n_{\uparrow',\downarrow'}(x,y)$. Thus, for each dressed
spin we readily obtained the temperature, total number $N_{\rm
T\uparrow',\rm T\downarrow'}$, and condensate densities $n_{\uparrow',\downarrow'}(x,y)$.

To analyze the miscibility from the TOF images where a
Stern-Gerlach gradient separated individual spin states, we
recentered the distributions to obtain $n_{\uparrow'}(x,y)$ and
$n_{\downarrow'}(x,y)$. This took into account the
displacement due to the Stern-Gerlach gradient and the nonzero
velocities $\hbar{\mathpzc k}_x/m$ of each spin state (after the
adiabatic mapping). The two origins were determined by the
following: we loaded the dressed states at a desired coupling
$\Omega$ but with detuning $\delta$ chosen to put all atoms in
either $\ket{\downarrow'}$ or $\ket{\uparrow'}$. Since
$q_{\uparrow,\downarrow}=\mp (1-\Omega^2/32E_L^2)\kl$ (see
Fig.~\ref{setup}c), these velocities $\hbar{\mathpzc
k}_x/m=\hbar(q_{\uparrow}+k_L)/m,\hbar(q_{\downarrow}-k_L)/m$ depend
slightly on $\Omega$, and our technique to determine the
distributions' origin accounts for this effect.

\subsection{Calibration of Raman Coupling}
Both Raman lasers were derived from the same Ti:Sapphire laser at
$\lambda\approx804.1$~nm, and were offset from each other by a pair
of AOMs driven by two phase locked frequency synthesizers near
$80\MHz$. We calibrated the Raman coupling strength $\Omega$ by
fitting the three-level Rabi oscillations between the $m_F=-1,0,\
{\rm and}\ +1$ states driven by the Raman coupling to the expected
behavior.

\end{methods}

\end{document}